\documentclass[twocolumn,apj]{emulateapj}
\usepackage{hyperref}

\usepackage{amsmath}

\begin{document}

\title{NuSTAR observations of high altitude post-flare loops one day after the flare}
\title{High-altitude Post-flare Loops seen by NuSTAR one Day after Flare Onset}
\title{High-altitude Post-flare Loops seen by NuSTAR one Day after Onset of a Solar Flare}
\title{High-altitude Loops one Day after Onset of a Solar Flare as seen by NuSTAR}
\title{NuSTAR Observations of High-altitude Solar Loops one Day after Flare Onset}
\title{NuSTAR Observations of High-altitude Loops one Day after the Onset of a Solar Flare}
\title{Evidence of significant energy input in the late phase of a solar flare from NuSTAR X-ray observations}

\author{Matej Kuhar\altaffilmark{1,2}, S\"am Krucker\altaffilmark{1,3}, Iain G. Hannah\altaffilmark{4}, Lindsay Glesener\altaffilmark{5}, Pascal Saint-Hilaire\altaffilmark{3}, Brian W. Grefenstette\altaffilmark{6}, Hugh S. Hudson\altaffilmark{3, 7}, Stephen M. White\altaffilmark{8}, David M. Smith\altaffilmark{9}, Andrew J. Marsh\altaffilmark{9}, Paul J. Wright\altaffilmark{4}, Steven E.~Boggs\altaffilmark{3}, Finn E. Christensen\altaffilmark{10},William W. Craig\altaffilmark{3, 11}, Charles J. Hailey\altaffilmark{12}, Fiona A.~Harrison\altaffilmark{6}, Daniel Stern\altaffilmark{13}, William W. Zhang\altaffilmark{14}}
\altaffiltext{1}{University of Applied Sciences and Arts Northwestern Switzerland, Bahnhofstrasse 6, 5210 Windisch, Switzerland}
\altaffiltext{2}{Institute for Particle Physics, ETH Z\"{u}rich, 8093 Z\"{u}rich, Switzerland}
\altaffiltext{3}{Space Sciences Laboratory, University of California, Berkeley, CA 94720-7450, USA}
\altaffiltext{4}{SUPA School of Physics \& Astronomy, University of Glasgow, Glasgow G12 8QQ, UK}
\altaffiltext{5}{School of Physics and Astronomy, University of Minnesota - Twin Cities , Minneapolis, MN 55455, USA}
\altaffiltext{6}{Cahill Center for Astrophysics, 1216 E. California Blvd, California Institute of Technology, Pasadena, CA 91125, USA}
\altaffiltext{7}{School of Physics and Astronomy, University of Glasgow, Glasgow G12 8QQ, UK}
\altaffiltext{8}{Air Force Research Laboratory, Albuquerque, NM, USA}
\altaffiltext{9}{Physics Department and Santa Cruz Institute for Particle Physics, University of California, Santa Cruz, 1156 High Street, Santa Cruz, CA 95064, USA}
\altaffiltext{10}{DTU Space, National Space Institute, Technical University of Denmark, Elektrovej 327, DK-2800 Lyngby, Denmark}  
\altaffiltext{11}{Lawrence Livermore National Laboratory, Livermore, CA 94550, USA} 
\altaffiltext{12}{Columbia Astrophysics Laboratory, Columbia University, New York, NY 10027, USA} 
\altaffiltext{13}{Jet Propulsion Laboratory, California Institute of Technology, 4800 Oak Grove Drive, Pasadena, CA 91109, USA} 
\altaffiltext{14}{NASA Goddard Space Flight Center, Greenbelt, MD 20771, USA} 

%\and

\begin{abstract}
We present observations of the occulted active region AR12222  during the third {\em NuSTAR} solar campaign on 2014 December 11, with concurrent {\em SDO/}AIA and {\em FOXSI-2} sounding rocket observations. The active region produced a medium size solar flare one day before the observations, at $\sim18$UT on 2014 December 10, with  the post-flare loops still visible at the time of {\em NuSTAR} observations. The time evolution of the source emission in the {\em SDO/}AIA $335\textrm{\AA}$ channel reveals the characteristics of an extreme-ultraviolet late phase event, caused by the continuous formation of new post-flare loops that arch higher and higher in the solar corona. The spectral fitting of {\em NuSTAR} observations yields an isothermal source, with temperature  $3.8-4.6$ MK, emission measure $0.3-1.8 \times 10^{46}\textrm{ cm}^{-3}$, and density estimated at $2.5-6.0 \times 10^8 \textrm{ cm}^{-3}$. The observed AIA fluxes are consistent with the derived {\em NuSTAR} temperature range, favoring temperature values in the range $4.0-4.3$ MK.  By examining the post-flare loops' cooling times and energy content, we estimate that at least 12 sets of post-flare loops were formed and subsequently cooled between the onset of the flare and {\em NuSTAR} observations, with their total thermal energy content an order of magnitude larger than the energy content at flare peak time. This indicates that the standard approach of using only the flare peak time to derive the total thermal energy content of a flare can lead to a large underestimation of its value.
 \end{abstract}
 
\keywords{Sun: flares --- Sun: particle emission --- Sun: X-rays}

\section{Introduction}
\begin{figure*}[hbtp]
\centering
\includegraphics[scale=0.71]{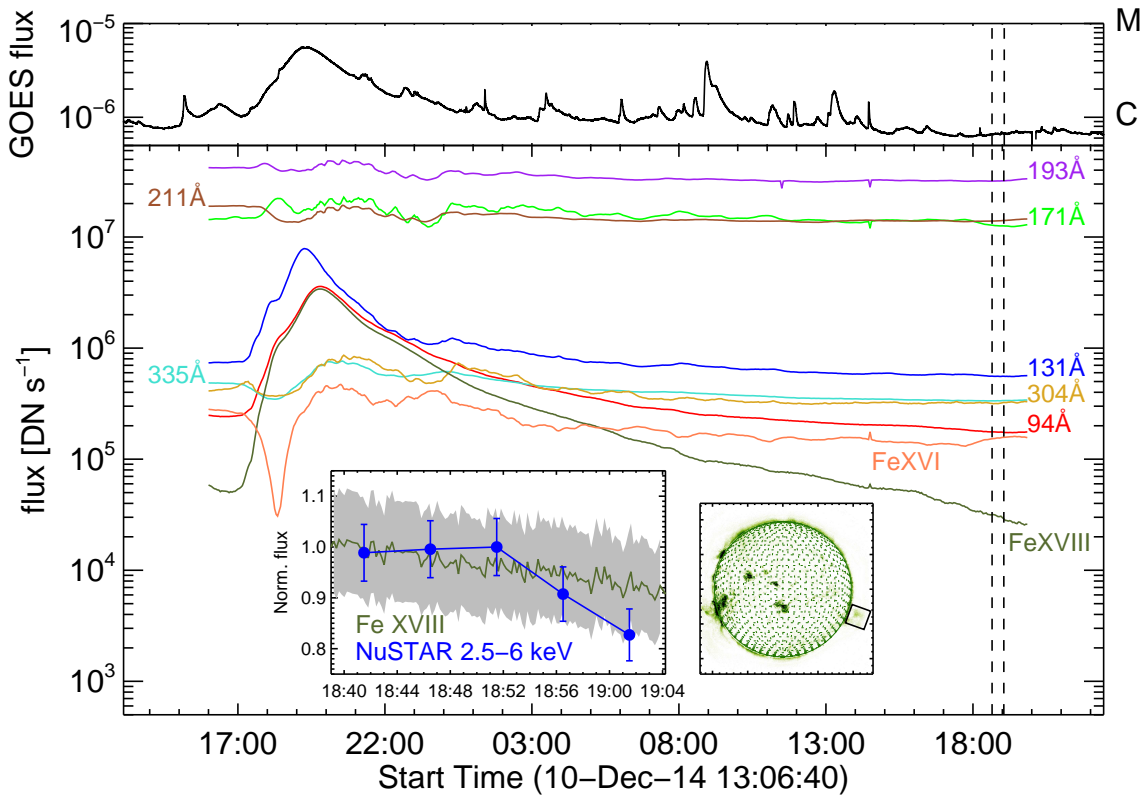}
\caption{Time profiles of GOES, 7 EUV channels of {\em SDO/AIA}, Fe {\sc xvi}, Fe {\sc xviii} and {\em NuSTAR} FPMB fluxes from the flaring area above the west limb as marked by the black box in the Fe {\sc xviii} map in the inset. Vertical dashed lines represent the time range of {\em NuSTAR} observations of the occulted active region AR12222. The inner plot shows the normalized Fe {\sc xviii} (olive line) and {\em NuSTAR} fluxes (blue dots) during the observations.} The grey shaded area represents an assumed uncertainty of 10\% in the Fe {\sc xviii} flux.
\label{fig1}
\end{figure*}

The \textit{Nuclear Spectroscopic Telescope ARray} ({\em NuSTAR}) is a focusing hard-X ray (HXR) telescope operating in the energy range from 3 to 79 keV \citep{Harrison13}. While primarily designed to observe far, faint astrophysical sources such as active galactic nuclei (AGN), black holes and supernova remnants, it is also capable of observing the Sun. With its focusing optics system, it can directly observe HXRs from previously undetected sources on the Sun due to its ten-times higher effective area and orders of magnitude reduced background when compared to state-of-the art solar HXR instruments such as \textit{Reuven Ramaty High Energy Solar Spectroscopic Imager} ({\em RHESSI}, \citealt{Lin02}). However, because it is optimized for observations of astrophyscial objects, {\em NuSTAR} experiences some technical challenges when observing the Sun; these include \textit{ghost-rays} and low throughput. Ghost-rays are unfocused, single-bounced photons (in contrast to properly focused  photons which reflect twice off the Wolter-I mirrors) coming from sources outside the field-of-view \citep{Madsen15}. The throughput  of {\em NuSTAR}'s focal plane detector electronics, with a maximum of 400 counts per second per telescope, can effectively diminish the hard X-ray sensitivity in the presence of extremely bright sources \citep{Grefenstette16}, making detections of fainter spectral components (such as a non-thermal component) difficult.
\par Despite these challenges, {\em NuSTAR} has begun to provide critical new observations of faint X-ray sources on the Sun  \citep{Hannah16} and giving us new insights into the coronal heating problem and particle energization in solar flares. In that respect, occulted active regions are priority targets in the planning of {\em NuSTAR} observations. With the brightest emission from the footpoints and low corona hidden, NuSTAR can search for faint coronal signature of heated material and particle acceleration. In order to maximize {\em NuSTAR} livetime and minimize ghost-rays during these observations, they should be carried out during low-activity periods (preferably with no other active sources on disk).
\par In this paper, we analyze the occulted active region AR12222 which produced a C5.9 \textit{GOES} (\textit{Geostationary Operational Environmental Satellite})  class flare $\sim24$ hours before {\em NuSTAR} observations. AR12222 was observed in the third {\em NuSTAR} solar campaign on 2014 December 11. The active region was also observed by \textit{Solar TErrestrial RElations Observatory} ({\em STEREO}), Atmospheric Imaging Assembly on \textit{Solar Dynamics Observatory} ({\em SDO/}AIA) and the second launch of \textit{Focusing Optics X-ray Solar Imager} ({\em FOXSI-2}) sounding rocket. The goal of this paper is to analyze the  time evolution of the X-ray and extreme-ultraviolet (EUV)  emission of the observed source above the solar limb in the context of the flare evolution scenario proposed by \cite{Woods11} and \cite{Woods14}. In these papers, the authors argue that flares may have four distinct phases in their evolution: (1) impulsive phase (best seen in HXRs), (2) gradual phase seen in SXR/EUV from the post-flare loops, (3) coronal dimming, best seen in the $171\textrm{\AA}$ line and (4) an EUV-late phase, best seen as a second peak in the $335 \textrm{\AA}$ line  few (up to 6)  hours after the flare onset. The explanation of the EUV late-phase emission lies in the formation of subsequent flare loops, overlying the original flare loops, which result from the reconnection of magnetic fields higher than those that reconnected during the flare's impulsive phase. Similar observations of ``giant post-flare loops" and ``giant arches" can be found in \cite{MacCombie79}, \cite{Svestka82}, \cite{ Svestka84}, \cite{Svestka95}, \cite{Farnik96}, \cite{Parenti10} and \cite{Matthew15}, among others; a theoretical model of the subsequent magnetic reconnections (and its successful description of the flare SOL1973-07-29T13)  is given in \cite{Kopp1984}. More recently, \cite{Liu13} proposed that the subsequent loop system(s) is produced by magnetic reconnection of the overlying active region magnetic field lines and the loop arcade produced by the flare, adding more complexity to the theoretical description of these events.
\par This paper is structured as follows. In Section 2 we give an overview of {\em NuSTAR}, {\em SDO/}AIA, {\em STEREO} and {\em FOXSI-2} observations of AR12222. We present the results of {\em NuSTAR} spectroscopy in Section 3, along with the comparison of {\em NuSTAR} derived parameters with observations in other wavelengths. The discussion of the results, as well as possible future studies, is presented in Section 4. 

\section{Observations}

\begin{figure*}[hbtp]
\centering
\includegraphics[scale=0.8]{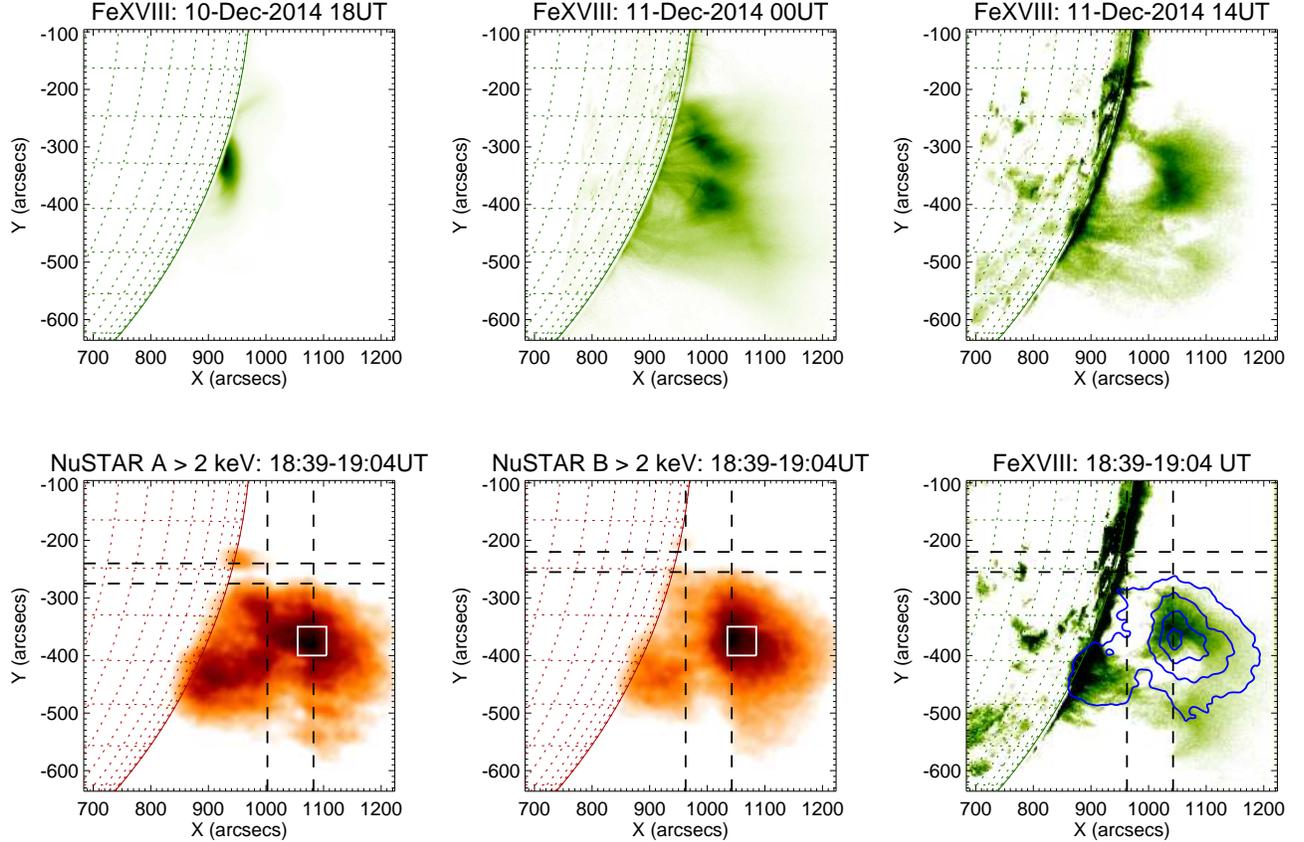}
  \caption{\textit{Upper row}: Fe {\sc xviii} maps of the flare onset (\textit{left} panel), post-flare loops 6 hours after the flare (\textit{central} panel), and the remaining loops 20 hours after the flare (\textit{right} panel). \textit{Bottom row:} 25-minute integrated {\em NuSTAR} FPMA (\textit{left} panel) and FPMB  (\textit{central} panel)  and AIA Fe {\sc xviii}  (\textit{right} panel) maps; the latter includes the 30\%, 50\%, 70\% and 90\% {\em NuSTAR} contours in blue. Dashed lines denote area affected by the NuSTAR chip gap during this observation. The white box is the region chosen for the spectral analysis.}
\label{fig2}
\end{figure*}

The data presented in this paper come from the third set of solar observations with {\em NuSTAR}, which were carried out on 2014 December 11. The observations consisted of observations of the north pole region (quiet Sun observations) and the solar limb (from 18:39:00 to 19:04:00 UT) that is discussed in this paper. 
\par The target for the limb pointing and of this study is the active region AR12222, located $\sim35$ degrees behind the south-west solar limb at the time of the {\em NuSTAR} observations. AR12222 produced a {\em GOES} C5.9 flare one day before the {\em NuSTAR} observations, at 18UT on 2014 December 10. Figure 1 presents the time evolution of the {\em GOES} flux, the 7 {\em SDO/}AIA EUV channels and the AIA-derived Fe {\sc xvi} and Fe {\sc xviii} fluxes from flare onset until more than a day later. The {\em NuSTAR} observing period is indicated with vertical dashed lines. Smaller spikes in the {\em GOES} curve between the flare and {\em NuSTAR} observations represent various fainter flares coming from other active regions (AR12233, AR12230, AR12235) on the solar disk. Due to the high occultation, the estimate of the {\em GOES} class as given above of the flare SOL2014-12-10T18 is a severe lower limit of the actual {\em GOES} class. The {\em STEREO} satellites can generally be used to give a prediction of the actual {\em GOES} class as they view the Sun from a different angle \citep{Nitta13}. Even though {\em STEREO A} was at the right location at an angle of $\sim175^{\circ}$ with respect to the Earth, it was not observing during the main and gradual phases of the flare; therefore, we cannot give an accurate {\em GOES} class estimate for this flare.  
\par The time evolution of fluxes in different AIA channels reveals two main characteristics of an EUV late-phase event, as described in \cite{Woods11} and \cite{Woods14}: a second (in this case weaker) peak in the $335\textrm{\AA}$ line a few hours after the flare, and  coronal dimming in the $171\textrm{\AA}$ line with the local minimum $\sim$5 hours after the flare. As previously noted \citep[e.g., ][]{Stewart74, Rust76, Hudson98, Zarro99, Howard04, McIntosh07}, there is a strong correlation between coronal dimming and coronal mass ejection (CME) events; indeed, a strong CME with the velocity of $\sim$1000 km s$^{-1}$ was associated with this nominally C-class flare\footnote{Data taken from the LASCO CME Catalog: \url{http://cdaw.gsfc.nasa.gov/CME_list/}.}.

\par The olive curve in Figure 1 presents the time evolution of the Fe {\sc xviii} line flux. An estimate of the emission in the Fe {\sc xviii} line can be constructed from the 94$\textrm{\AA}$ line, by subtracting the lower temperature responses from the 171$\textrm{\AA}$, 193$\textrm{\AA}$ and/or 211$\textrm{\AA}$ channels \citep[see][]{delZanna13, Reale11, Testa12, Warren12}. In obtaining the Fe {\sc xviii} flux, we followed the approach of \cite{delZanna13}, using the formula
\begin{equation}
F(\textrm{Fe {\sc xviii}}) \approx F(94 \textrm{\AA})-F(211\textrm{\AA})/120-F(171\textrm{\AA})/450,
\end{equation}
where $F(\textrm{Fe {\sc xviii}})$ is the Fe {\sc xviii} flux, $F(94 \textrm{\AA}), F(211 \textrm{\AA})$ and $F(171 \textrm{\AA})$ are the fluxes in the $94 \textrm{\AA}, 211 \textrm{\AA} \textrm{ and } 171 \textrm{\AA}$ channels, respectively. 
The Fe {\sc xviii} line has a strong response in the temperature range from $\sim$3 to $\sim$10 MK, with the peak  around 6.5 MK.  The Fe {\sc xviii} time evolution shows a strong peak due to the flare, with a long decay phase lasting past the {\em NuSTAR} observations.
\par Similar to the Fe {\sc xviii} line, a lower-temperature Fe {\sc xvi} line  can be constructed from the $335 \textrm{\AA}$ and $171 \textrm{\AA}$ lines \citep{delZanna13}: 
\begin{equation}
F(\textrm{Fe {\sc xvi}}) \approx F(335\textrm{\AA})-F(171\textrm{\AA})/70.
\end{equation}
Similar to Fe {\sc xviii}, the above formula is just an approximation of the Fe {\sc xvi} flux. The Fe {\sc xvi} line has a temperature response of similar shape to the Fe {\sc xviii} line, with its peak at a lower temperature of $\sim2.5$ MK. The time evolution of the Fe {\sc xvi} flux is also shown in Figure \ref{fig1}. It is characterized by a strong dip followed by the initial rise, soon after which a decrease is observed, due to the fact that the flare becomes weaker. After $\sim8$UT on 2014 December 11, the time evolution of  Fe {\sc xvi} flux is determined by fore- and background emission along the line-of-sight, making the post-flare loops no longer observable in this line.
\par The evolution of 5-minute integrated {\em NuSTAR} fluxes (blue dots) and Fe {\sc xviii} fluxes (olive line) is given in the inset of Figure \ref{fig1}. The {\em NuSTAR} and Fe {\sc xviii} time evolutions show similar behaviour, with the (slow) decay rate of the two agreeing within the error bars and the only difference being the steeper decay of {\em NuSTAR} flux towards the end of the observation, which is likely an instrumental effect. The {\em NuSTAR} focal plane consists of a $2 \times 2$ array of CdZnTe detectors, which are divided into quadrants by a chip gap \citep{Harrison13}. As the telescope pointing drifted slowly during the observations, the gap covered part of the area used for calculating the flux. Therefore, it is probable that the steeper decay of the {\em NuSTAR} emission towards the end of the observation is not due to solar variability, but rather a consequence of the telescope drift. This might also have some effect on the determination of the temperature and emission measure of the source, which will be discussed in the following sections.

\par Due to the slow decay of Fe {\sc xviii} emission, we were able to make Fe {\sc xviii} images  even at the time of {\em NuSTAR} observations one day after the flare onset (see Figure \ref{fig2}). The upper row presents the Fe {\sc xviii} maps of the flare onset, the post-flare loops 6 hours after the flare, and the remaining  features 20 hours after the flare. Left and central panels in the bottom row present  25-minute integrated {\em NuSTAR} images above 2 keV from focal plane modules A (FPMA) and B (FPMB). Dashed lines denote the area covered by the gap during the observations that is further enlarged due to the drift of the telescope. As the drift was dominantly along the x-direction (45 arcsecs in total) and negligible in the y-direction, the area affected by the gap is much larger in the x-direction.  The region of interest for the analysis that will be presented in the next section, with an area of $50'' \times 50'' = 2500 \textrm{ arcsec}^2$, is marked by the white box. The last image in the bottom right corner is the 25-minute integrated (same time range as {\em NuSTAR}) Fe {\sc xviii} map of the source together with the  30, 50, 70 and 90\% contours of {\em NuSTAR} emission in blue. As the uncertainty in {\em NuSTAR} absolute pointing accuracy  is relatively large (see \citealt{Hannah16}, \citealt{Grefenstette16}), the {\em NuSTAR} image was shifted by $-100''$ and $25''$ in the x  and y directions, respectively, in order to match the Fe {\sc xviii} source location. {\em NuSTAR} and Fe {\sc xviii} maps show the same sources, such as the top parts of the coronal loops, and the high emission source above them \citep{MacCombie79}. 

\begin{figure}[hbtp]
 \centering
 \includegraphics[scale=0.49]{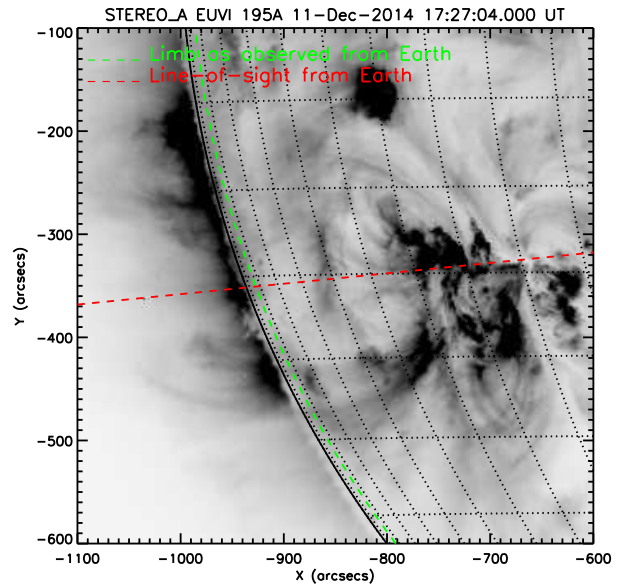}
\caption{{\em STEREO A} $195 \textrm{\AA}$ image of active region AR12222 an hour before the {\em NuSTAR} observations. The orange line presents the solar limb as viewed from the Earth, while the red line is the line-of-sight from the Earth through the {\em NuSTAR} source.}
  \label{fig3}
\end{figure}

\par In Figure \ref{fig3} we present the {\em STEREO A} image of active region AR12222 an hour before the {\em NuSTAR} observation. The orange line shows the solar limb as viewed from the Earth, while the red line is a projection of the line-of-sight from the Earth to the {\em NuSTAR} source, passing right above AR12222 located at $\sim$[$-730'', -330''$] in the {\em STEREO A}  $195\textrm{\AA}$ image. The {\em NuSTAR} source is not evident in this image as the $195\textrm{\AA}$ channel is sensitive only to lower temperatures. From {\em STEREO} images, it is possible to calculate the height of the post-flare loops, defined as the distance between AR12222 and the mid-point of the line that minimizes the distance between the Earth-Sun line-of-sight and the radial extension above the active region. We estimate this height to be $\sim300''$. If we assume the height of the original loops at the flare onset to be $50''$ (as there are no {\em STEREO} observations of this active region immediately after the flare, we assume this height as a common value for ordinary flares), this yields a radial velocity of  $ \sim 2 \textrm{ km s}^{-1}$ when averaged over the whole day. This is similar to typical speeds of rising post-flare loops very late in an event \cite[e.g., ][]{MacCombie79, Gallagher02}, giving further evidence that the {\em NuSTAR} source is indeed associated with the flare that occurred a day earlier.\\

\begin{figure*}[hbtp]
\centering
\includegraphics[scale=0.8]{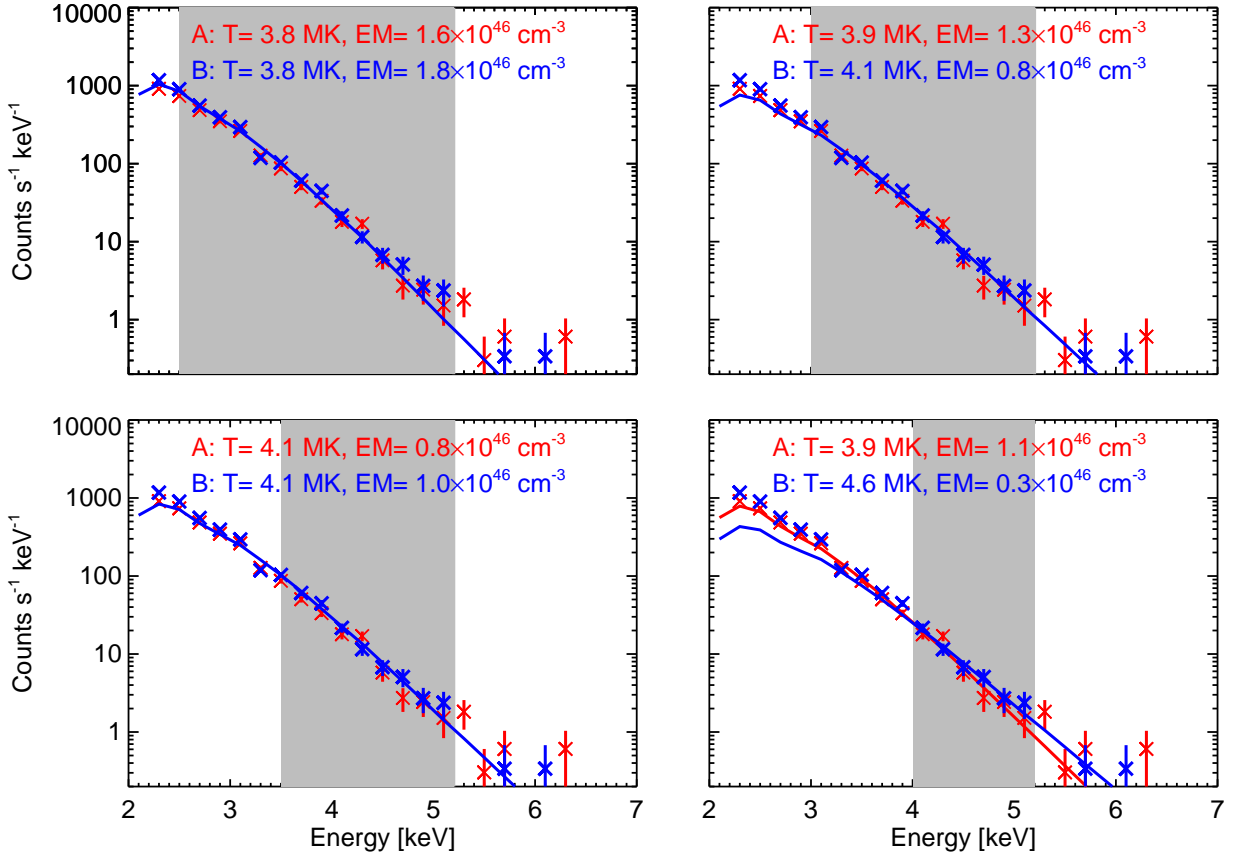}
\caption{{\em NuSTAR} count spectra for FPMA (red) and FPMB (blue) integrated over the whole observation time range ($18:39-19:04$), together with isothermal fits for different energy ranges: $2.5-5.2$ keV (\textit{upper left} ), $3.0-5.2$ keV (\textit{upper right}), $3.5-5.2$ keV (\textit{lower left}) and $4.0-5.2$ keV (\textit{lower right}). Energy ranges for spectral fitting are shown with grey shaded areas. The best-fit values of temperature and emission measure for individual focal plane modules can be found on the top of each graph.}
\label{fig4}
\end{figure*}

\begin{figure*}[hbtp]
\centering
\includegraphics[scale=0.71]{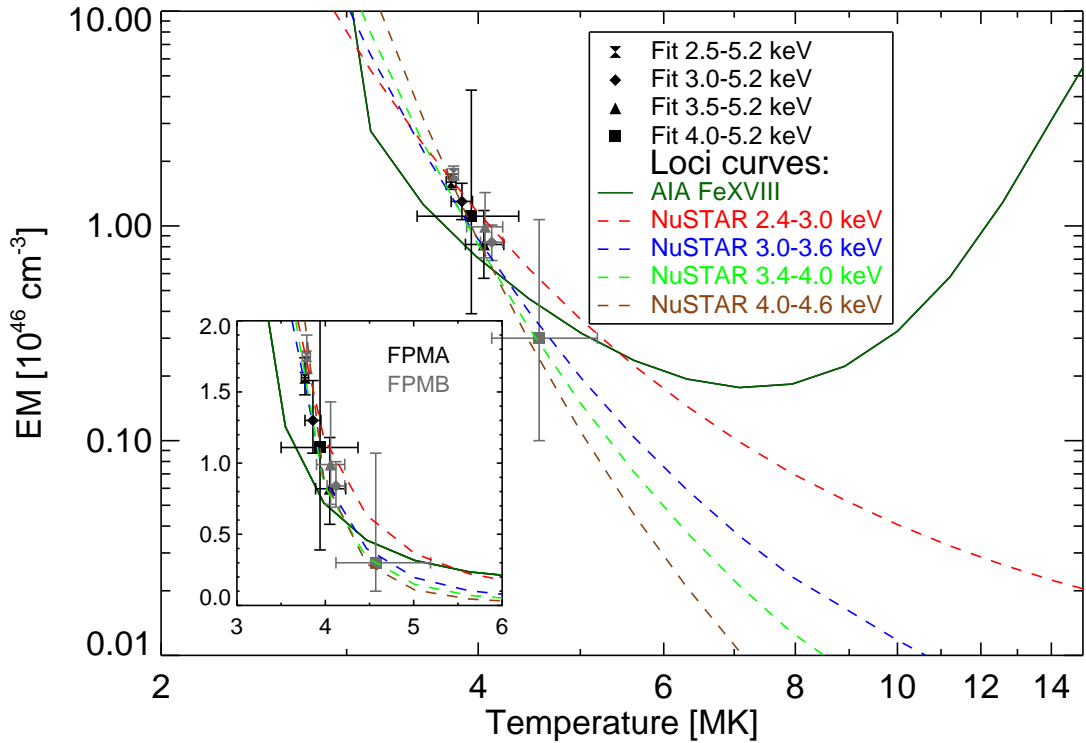}
\caption{Comparison of Fe {\sc xviii} and {\em NuSTAR} loci curves. Temperatures and emission measures from the fits in different energy ranges are marked by symbols with corresponding confidence ranges in black (FPMA) and grey (FPMB). The inset represents the magnification of the given plot in the temperature range  $3-6$ MK, on a linear scale.}
\label{fig5}
\end{figure*}

\section{Analysis of the high coronal source}
\subsection{Spectral fitting}
We fitted the {\em NuSTAR} count spectrum inside the region of interest from Figure \ref{fig2} separately for FPMA and FPMB, following the approach of Hannah et al. 2016, using SolarSoft/OSPEX\footnote{ \url{http://hesperia.gsfc.nasa.gov/ssw/packages/spex/doc/}.}. The counts were binned with 0.2 keV energy resolution, while the integration time was 25 minutes (full {\em NuSTAR} observing time of the active region).  As the livetime was around 1\% during the whole observation period, this is roughly equal to 15 seconds of exposure at full livetime. In order to investigate the influence of the adopted energy range on the fitted temperature and emission measure, we fitted CHIANTI 7.1 isothermal models \cite[][]{dere97, Landi13} to our data for different energy ranges: 2.5--5.2, 3.0--5.2, 3.5--5.2, 4.0--5.2 keV. These fits are presented in Figure \ref{fig4}. The lower limit of 2.5 keV was chosen as the lowest energy for which the calibration is still completely understood and reliable \citep{Grefenstette16}, while the upper limit of 5.2 keV was chosen as the highest energy with a significant number of counts ($>3$ counts per bin). Both focal plane modules give consistent results, with temperature  $3.8-4.6 \textrm{ MK}$ and emission measure $0.3 \times 10^{46}  \textrm{ cm}^{-3} - 1.8 \times 10^{46} \textrm{ cm}^{-3}$, depending on the lower limit of the energy range used in the fitting. The temperature gets higher and the emission measure gets lower as we go to higher energies. The 67\% confidence ranges of temperature and emission measure were calculated using the standard Monte Carlo procedure in OSPEX and  are given in Table 1 together with the best-fit values. A point to note is that our region of interest is located very close to the gap between the detectors, which leads to fewer counts, especially in later phases of the integration interval. The reason for this is the slow drift of the spacecraft pointing with time, resulting in covering a part of the region of interest by the gap. The missing counts could lead to an underestimation of the emission measure, but do not change the value of the determined temperature (as it is determined by the slope in the counts spectrum).  A single temperature component is enough to fit the observations, similar to the results of \cite{Hannah16}. We determine the density of the source to be (assuming a volume of $50\times50\times50$ arcsec$^3$) in the range $2.5-6.0 \times 10^8 \textrm{cm}^{-3}$ (roughly $10-100$ times the density of the quiet Sun corona at this height; see e.g., \citealt{Withbroe88}), suggesting the density of late-phase loops to be significantly higher than that of the quiet Sun corona.

\begin{center}
\begin{table}
\centering
\label{table1}
\begin{tabular}{c| c | c|c | c}
\multicolumn{5}{c}{FPMA} \\
\hline   
\hline   

  Energy range [keV] & $2.5-5.2$        & $3.0-5.2 $& $3.5-5.2$ & $4.0-5.2$          \\
\hline 
   & & & &\\
T [MK]  &  $3.77\substack{+0.04 \\ -0.04}$& $3.86\substack{+0.09 \\ -0.09}$  &$4.05\substack{+0.18 \\ -0.16}$& $3.94\substack{+0.43 \\ -0.44}$ \\
& & \\
EM [$10^{46} \textrm{cm}^{-3}$] &  $1.60\substack{+0.14 \\ -0.12}$& $1.30\substack{+0.28 \\ -0.23}$  &$0.82\substack{+0.36 \\ -0.25}$& $1.11\substack{+3.18 \\ -0.72}$  \\\\

\multicolumn{5}{c}{FPMB} \\
\hline   
\hline   

  Energy range [keV] & $2.5-5.2$        & $3.0-5.2 $& $3.5-5.2$ & $4.0-5.2$          \\
\hline 
   & & & &\\
T [MK]  &  $3.79\substack{+0.04 \\ -0.05}$& $4.12\substack{+0.10 \\ -0.10}$  &$4.06\substack{+0.16 \\ -0.16}$& $4.57\substack{+0.62 \\ -0.45}$ \\
& & \\
EM [$10^{46} \textrm{cm}^{-3}$] &  $1.75\substack{+0.15 \\ -0.12}$& $0.84\substack{+0.17 \\ -0.15}$  &$0.99\substack{+0.44 \\ -0.28}$& $0.30\substack{+0.77 \\ -0.20}$  \\\\

\end{tabular}
\caption{Best-fit values of temperature and emission measure and their 67\% confidence ranges.}
\end{table}
\end{center}

\subsection{Comparison of NuSTAR to SDO/AIA}
\subsubsection{Comparison to Fe {\sc xviii}}

\par In order to investigate the extent of the agreement between {\em NuSTAR} and Fe {\sc xviii} sources, we compare the Fe {\sc xviii} loci curve with the {\em NuSTAR} loci curves in different energy channels. For reference, the results of {\em NuSTAR} spectral fitting from the previous section for both focal plane modules are presented  in Figure \ref{fig5} with different symbols for different energy ranges, together with the Fe {\sc xviii} and {\em NuSTAR} loci curves. The Fe {\sc xviii} loci curve is extracted from the temperature response functions \citep{Boerner14} and the observed fluxes using the following formula
\begin{equation}
EM=\frac{F \cdot S}{R(T)},
\label{eq2}
\end{equation}
where $EM$ is the emission measure [cm$^{-3}$], $F$ is the flux $[\textrm{DN s}^{-1} \textrm{pix}^{-1}]	$, $S$ is the area of the region [cm$^2$] and $R(T)$ is the temperature response function of the Fe {\sc xviii} line $[\textrm{DNcm}^{5} \textrm{s}^{-1} \textrm{pix}^{-1}]	$. The {\em NuSTAR} loci curves are extracted in a similar way from the {\em NuSTAR} temperature response function, determined by folding the generated photon spectra for different temperatures through the {\em NuSTAR} response matrix. The good agreement of our results is best seen in the inset of Figure \ref{fig5}, where we plot the loci curves and the determined $EM-T$ pairs on linear scale. The intersection of the Fe {\sc xviii} loci curve with the {\em NuSTAR} loci curves in the temperature range $4.0-4.3$ MK is consistent with the $EM-T$ pairs shown in Figure \ref{fig4}, except for the fit including the lowest energies. A part of these low energy counts might originate from cooler post-flare loops, which will also be discussed in more detail in the next sections.
\begin{figure}[hbtp]
\centering
\includegraphics[scale=0.40]{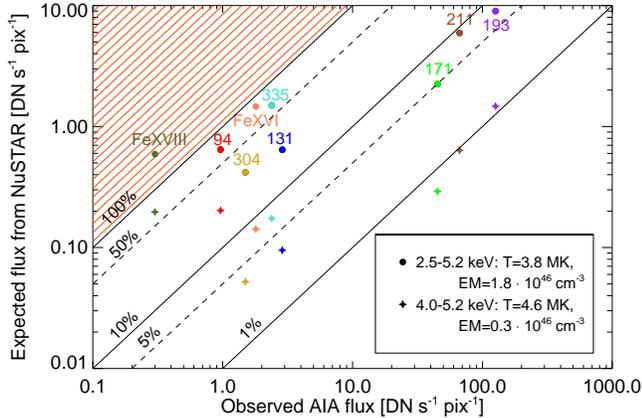}
\caption{Comparison of expected and observed fluxes for 7 AIA channels and the derived Fe {\sc xvi} and Fe {\sc xviii} channels, for the two extreme pairs of temperature and emission measure (values given in the legend) from fits in Figure \ref{fig4}. The diagonal lines denote 1, 5, 10, 50 and 100\% ratios of the expected AIA fluxes from the NuSTAR source and the observed AIA fluxes. The red lines denote the (forbidden) area where the predicted AIA flux from the {\em NuSTAR} source is larger than the total observed AIA flux.}
\label{fig6}
\end{figure}

\subsubsection{Comparison to other AIA channels}
\par It is also possible to investigate the results of {\em NuSTAR} fitting to other AIA channels by calculating the expected  count rates in different AIA channels from the source with the emission measure and temperature as given by {\em NuSTAR}, and compare them to the observed fluxes in AIA maps. The difficulty of this comparison is that the fraction of the cold background emission (in the temperature range below $\sim3$ MK) in these channels is unknown and non-removable. This is not an issue for  the derived Fe {\sc xviii} channel, which is not sensitive to this cooler plasma. The expected AIA fluxes are calculated by inverting Equation \ref{eq2}. This is a {\em NuSTAR}-predicted AIA flux coming from the {\em NuSTAR} source alone, without any additional contribution from the cooler plasma. The comparison between {\em NuSTAR}-predicted and observed fluxes is presented in Figure \ref{fig6}. The circles are the predicted fluxes for {\em NuSTAR} spectral fitting in the range $2.5-5.2$ keV, and the stars for $4.0-5.2$ keV. We use the fitted values of FPMB in both ranges, as they represent the two extreme $T-EM$ fits.  The full and dashed lines represent 1, 5, 10, 50 and 100\% ratios of {\em NuSTAR}-predicted and observed fluxes in different AIA channels. The area where the predicted AIA flux from the {\em NuSTAR} source is larger than the total observed flux is shown with the red lines. If the {\em NuSTAR} -predicted flux for a given AIA  channel is close to the observed flux (e.g., region between 50\% and 100\% lines in the plot), the emission in that AIA channel is dominated by the same plasma that {\em NuSTAR} observes. Unsurprisingly, this is best achieved for the $94\textrm{\AA}$ channel and, consequently, the Fe {\sc xviii} channel. For the first $T-EM$ fit, the  {\em NuSTAR}-predicted flux for the Fe {\sc xviii} channel is greater than the observed flux. This result indicates that a single temperature fit is not enough to fit the observations at the lowest energies, as some of the low-energy counts are produced by a lower temperature plasma. The ratio for the Fe {\sc xviii} channel for the fit at higher energies (second $T-EM$ fit) lies in the range between 50\% and 100\%, while the  $335 \textrm{\AA}$ channel and its derived Fe {\sc xvi} channel have ratios  in the range $5-10$ \%. These results are in agreement with the fact that the Fe {\sc xviii} source showed the same spatial features as the {\em NuSTAR} source, while we were not able to detect the Fe {\sc xvi} source.  Cooler lines at $171\textrm{\AA}, 211\textrm{\AA}$ and $193\textrm{\AA}$  have ratios of {\em NuSTAR}-predicted fluxes to the observed fluxes at a percent level, which is expected as these lines are sensitive to plasma cooler than NuSTAR can observe.

\par 
\subsection{Comparison of NuSTAR to FOXSI}
The {\em FOXSI} \citep{Krucker14} sounding rocket also uses direct focusing HXR optics, but is optimized especially for solar purposes. {\em FOXSI} has about one fifth of {\em NuSTAR}'s effective area with a higher spatial resolution (FWHM  of ~9 arcsec). The main difference for solar observations  between the two telescopes is the different low energy threshold. While {\em NuSTAR} detects photons down to $\sim$2 keV, the {\em FOXSI} entrance window intentionally blocks the large number of low energy photons, giving a typical peak in the count spectrum around 5 keV. The entrance window largely reduces the number of incoming photons, keeping the livetime high for the faint, higher-energy components. For example, a 25 minute observation by {\em NuSTAR} at 1\% livetime and five times the effective area is equal to a {\em FOXSI} observation of 75 s at full livetime. However, this also means that {\em FOXSI} is not sensitive to low temperature plasmas that are best seen below 4 keV.
\par The {\em FOXSI-2} rocket flew for a 6.5-minute observation interval during the {\em NuSTAR} solar pointing discussed here. {\em FOXSI-2} targeted AR12222 for 35.2 seconds, though 12 minutes after the {\em NuSTAR} observation finished. As the {\em NuSTAR}/AIA source has a slow time variation, the time difference between the observations is of minor importance, at least for the order-of-magnitude estimate discussed here. Using the temperature and emission measure derived from {\em NuSTAR} ($T=3.8$ MK and $EM=1.7 \times 10^{46} \textrm{ cm}^{-3}$), the expected {\em FOXSI} count rate is $\sim$1.6 counts for the {\em FOXSI-2}'s most sensitive  optics/detector pair D6. This value is computed above 5 keV and with the integration time of 35.2 seconds (integrating during the whole observation period). In total, 4 counts were observed by D6. This is a reasonable value given that the estimated non-solar background flux is 1.8 counts, while the expected count rate due to ghost-rays from sources outside of the FOV is unknown. Given the small-number statistics and the uncertainty of the ghost-ray background, the observed {\em FOXSI-2} measurement is consistent with the values expected for the plasma observed with {\em NuSTAR}, but does not provide any further diagnostics for this event.

\section{Dicussion and conclusions}
\par In this paper, we have presented the first observations of the EUV late-phase of a solar flare in X-rays with {\em NuSTAR}. {\em NuSTAR} has provided  unique opportunity to perform spectroscopy on X-rays from a coronal source a full day after the flare onset. With knowledge of the location of this faint source from {\em NuSTAR}, we were also able to find it in  Fe {\sc xviii} by eliminating the lower temperature response of the AIA $94\textrm{\AA}$ channel and integrating for 25 minutes (adding together 125 maps to obtain a higher signal-to-noise ratio). Here, {\em NuSTAR} played a crucial role in providing the information needed  for extracting the very faint signal which was far from evident in the $94\textrm{\AA}$ maps.
\par The fact that the post-flare loops have been observed so late in the flare evolution points to continuing energy input in the later phases of the solar flare evolution. To quantify this statement, we estimate the cooling times of subsequent post-flare loops and compare them to the flare duration. We follow the approach of \cite{Cargill95}, with the following formula for the cooling time of post-flare loops:
\begin{equation}
\tau_{cool}=2.35\cdot10^{-2} \cdot L^{5/6} \cdot n_e^{-1/6} \cdot T_e^{-1/6},
\label{eq3}
\end{equation}
where $\tau_{cool}$ [s] is the cooling time (the time needed for post-flare loops to cool down to $ \sim 10^5$ K) and $L$ [cm], $n_e$ [cm$^{-3}$] and $T_e$ [K] are loop length, density and temperature at the start time. The temperature estimate of the original post-flare loops from the {\em GOES} observations is 10.5 MK, while the emission measure is $5 \times 10^{48} \textrm{cm}^{-3}$. Even though the above estimates might only be a rough approximation because of the high occultation of the flare, we are anyway making only an approximate calculation of the cooling time. By assuming the length of the original post-flare loops to be $\sim50''$, we estimate the density to be $9 \times 10^{9} \textrm{cm}^{-3}$. This gives us a cooling time of $\sim$1 hour, indicating that the original post-flare loops are long-gone at the time of {\em NuSTAR} observations and that the additional heating took place during the evolution of the post-flare system. The most probable explanation is the previously mentioned scenario of subsequent magnetic reconnections, resulting in reconnected loops being produced higher and higher in the corona.
\par The above results are in agreement with original \textit{Skylab} and SMM results, and the recent observations of a large post-flare loop system between 2014 October $14-16$ by \cite{Matthew15}. They conclude that the giant late-phase arches are similar in structure to the ordinary post-flare loops, and formed by magnetic reconnection. Their reasoning follows  the work of \cite{Forbes00}, in which it is pointed out that the reconnection rate may not depend only on the magnetic field (in which case, it would decrease with height), but possibly on the local Alfven speed, which is proportional to $B/ \sqrt{\rho}$, where $B$ is the magnetic field strength and $\rho$ the density. So, if the density decreases sufficiently fast, the reconnection rate could remain constant out to $0.5R_{\sun}$ despite the decreasing magnetic field strength, and thus produce the giant post-flare loops analyzed by \cite{Matthew15} or in this study. 

\par From {\em NuSTAR} and {\em GOES} data, it is possible to estimate the additional energy input needed to form the subsequent, rising post-flare loops. The total thermal energy of the loop system is proportional to the density, temperature and volume \citep[e.g., ][]{Hannah08_2}:
\begin{equation}
E_{th}=3NkT=3k\cdot nVT,
\label{eq4}
\end{equation}
while $k$ is the Boltzmann constant. We have obtained all the above parameters for the original flare loops from {\em GOES} and for the post-flare loops a day after from {\em NuSTAR}. We estimate that the thermal energy content in {\em NuSTAR}-loops is 5\% of the thermal energy content of the original flare loops, indicating there is still significant energy release even a full day after the flare onset.  Next, by assuming linearity in the change of density, loop length and temperature over time (for simplicity), it is possible to calculate the change in cooling times of all the post-flare loops formed in between. Although the above assumption might not be accurate for all (or any) of the parameters, we are only interested in calculating an order of magnitude estimate here. The other assumption we use is that new loop systems are only produced when the old ones vanish. This assumption is in principle not valid as new systems are produced while the old ones persist, but it gives us an approximate lower limit on the total thermal energy content in all the loops systems. The sequence is as follows: original post flare loops vanish after $\sim1$ hour, and during this time density, temperature and volume change as well, and a new loop system with a different cooling time is produced. We calculate that this sequence repeats about 12 times during the 24 hours between the flare onset and {\em NuSTAR} observations, with the total energy content in those 12 cycles of reconnection and cooling estimated at a factor of $\sim13$ larger than the one released during the impulsive phase of the flare only. 
\par Previous estimates of the additional energy input during the decay phase of solar flares were derived using radiative losses at specific wavelength ranges. \cite{Woods11} calculate the total radiated energy in the EUV band during the late phase to be between 0.4 and 3.7 times the flare energy in the X-rays during the peak. \cite{Emslie12} conclude in their statistical study of 38 solar flares that, on average, the total energy radiated from hot SXR-emitting plasma exceeds the peak thermal energy content by a factor of $\sim3$. It is important to note that the above studies used non-overlapping wavelength ranges, thus missing the contribution to total energy content from the wavelength range of the other study (and the rest of the wavelength spectrum). Our results for a single event are consistent with these statistical studies, especially as we compare our value with statistical averages that miss significant energy contributions.
\par In summary, all results indicate that the impulsive energy release is only a fraction of the energy release in the late phase of the flare evolution, at least for events with clearly observable late phase emission. This statement calls for re-examining the approach of using just the peak energy content or the non-thermal emission during the impulsive phase of the flare as the estimate of the total energy content of the flare. In order to assess this in more detail, a statistical study of similar events should be carried out. However, {\em NuSTAR} is not a solar-dedicated observatory, and therefore the observations are few and sporadic, making statistical studies difficult. Additionally, it is most likely that faint signals such as presented in this study can only be observed when the flare (and the active region) is occulted or at least over the limb, as the emission from these kinds of coronal sources on the disk would likely be masked by the much stronger emission of the active region beneath. Nevertheless, a statistical search for {\em SDO/AIA} Fe {\sc xviii} sources in above-the-limb flares could give us new insights about the influence of the long-lasting decay phase on flare energetics. 

\acknowledgments

This work made use of data from the NuSTAR mission, a project led by the California Institute of Technology, managed by the Jet Propulsion Laboratory, and funded by NASA. We thank the NuSTAR Operations, Software and Calibration teams for support with the execution and analysis of these observations. This research made use of the NuSTAR Data Analysis Software (NuSTARDAS), jointly developed by the ASI Science Data Center (ASDC, Italy) and the California Institute of Technology (USA). M.K. and S.K. acknowledge funding from the Swiss National Science Foundation (200021-140308). Funding for this work was also provided under NASA grants NNX12AJ36G and NNX14AG07G. A.J.M.'s participation was supported by NASA Earth and Space Science Fellowship award NNX13AM41H. I.G.H. is supported by a Royal Society University Research Fellowship. P. J. W. is supported by an EPSRC-Royal Society fellowship engagement grant. FOXSI was funded by NASA LCAS grant NNX11AB75G.

\bibliographystyle{apj}
\bibliography{journals,whitelight}

\end{document}